\documentclass{amsart}
\begin{document}
\title{Approximate symmetries of the Harry Dym equation}
%
\author[ M. Nadjafikhah, P. Kabi-Nejad]{ Mehdi Nadjafikhah, Parastoo Kabi-Nejad}
\begin{abstract} In this paper, we derive the first order approximate symmetries for the Harry Dym equation by the method of approximate transformation groups proposed by Baikov, Gaszizov and Ibragimov \cite{1}, \cite{2}. Moreover, we investigate the structure of the Lie algebra of symmetries of the perturbed Harry Dym equation. We compute the one-dimensional optimal system of subalgeras as well as point out some approximately differential invarints with respect to the generators of Lie algebra and optimal system.
\end{abstract}
\maketitle
%
\section{Introduction}
The following nonlinear partial differential equation
\begin{eqnarray}\label{eq:1}
u_t=-\frac{1}{2}u^3u_{xxx}.
\end{eqnarray}
is known as the Harry Dym equation \cite{7}. This equation was obtained by Harry Dym and Martin Kruskal as an evolution equation solvable by a spectral problem based on the string equation instead of Schr\"{o}dinger equation. This result was reported in \cite{9} and rediscovered independently in \cite{16}, \cite{17}. The Harry Dym equation shares many of the properties typical of the soliton equations. It is a completely integrable equation which can be solved by inverse scattering transformation \cite{3}, \cite{18}, \cite{19}. It has a bi-Hamiltonian structure and an infinite number of conservation laws and infinitely many symmetries \cite{11}, \cite{12}.

In this paper, we analyze the perturbed Harry Dym equation 
\begin{eqnarray}\label{eq:22}
u_t+\frac{1}{2}u^3u_{xxx}+\varepsilon u_x=0,
\end{eqnarray}
where $\varepsilon$ is a small parameter, 
with a method which was first introduced by 
 Baikov, Gazizov and Ibragimov \cite{1}, \cite{2}. This method which is known as "approximate symmetry"
  is a combination of Lie group theory and perturbations. There is a second method which is also 
  known as "approximate symmetry" due to Fushchich anf Shtelen \cite{6} and later followed by Euler et al \cite{4}, 
  \cite{5}. For a comparison of these two methods, we refer the interested reader to the papers \cite{13}, \cite {15}.  
our paper is organized as follows: 
In section 2, we present some definitions and theorems in the theory of approximate symmetry. 
 In section 3, we obtain the  approximate symmetry of  
the perturbed Harry Dym equation.  In section 4, we discuss on the structure of its Lie algebra. In section 5, we construct 
the one-dimensional optimal system of subalgebras. In section 6, we compute some approximately differential invarints with respect 
to the generators of Lie algebra and optimal system.
In section 7, we summarize our results.  
\section{Notations and Definitions}
In this section, we will provide the background definitions and results in approximate symmetry that 
will be used along this paper. Much of it is stated as in \cite{8}.
If a function$f(x,\varepsilon)$ satisfies the condition
\begin{eqnarray}\label{eq:2}
\lim \dfrac{f(x,\varepsilon)}{\varepsilon^p}=0,
\end{eqnarray}
it is written $f(x,\varepsilon) = o(\varepsilon^p)$ and f is said to be oforder less than $\varepsilon^p$.
If
\begin{eqnarray}\label{eq:3}
f(x,\varepsilon)-g(x,\varepsilon)=o(\varepsilon^p),
\end{eqnarray}
the functions $f$ and $g$are said to be approximately equal (with an error $o(\varepsilon^p)$) and
written as 
\begin{eqnarray}\label{eq:4}
f(x,\varepsilon)=g(x,\varepsilon)+o(\varepsilon^p),
\end{eqnarray}
or or, briefly $f\approx g$  when there is no ambiguity.
The approximate equality defines an equivalence relation, and we join functions
into equivalence classes by letting $f(x,\varepsilon)$ and $g(x,\varepsilon)$ to be members of the same
class if and only if $f\approx g$.
Given a function $f(x,\varepsilon)$, let
\begin{eqnarray}\label{eq:5}
f_o(x) +\varepsilon f_l(x) +\cdots+\varepsilon^p f_p(x)
\end{eqnarray}
be the approximating polynomial of degree p in $\varepsilon$ obtained via the Taylor series expansion
of $f(x,\varepsilon)$  in powers of $\varepsilon$ about $\varepsilon= 0$.
Then any function $g\approx f$ (in particular, the function $f$ itself) has the form
\begin{eqnarray}\label{eq:6}
g(x,\varepsilon)=f_o(x) +\varepsilon f_l(x) +\cdots+\varepsilon^p f_p(x)+o(\varepsilon^p).
\end{eqnarray}
Consequently the expression (\ref{eq:5}) is called a canonical representative of the equivalence class of functions containing
$f$· Thus, the equivalence class of functions $g(x,\varepsilon)  f(x,\varepsilon)$ is determined by the
ordered set of $p+1$ functions $f_0(x), f_l(x),\cdots, f_p(x)$.
In the theory of approximate transformation groups, one considers ordered sets of
smooth vector-functions depending on $x$'s and a group parameter $a$:
\begin{eqnarray}\label{eq:7}
f_0(x,a), f_l(x,a),\cdots, f_p(x,a),
\end{eqnarray}
with coordinates
\begin{eqnarray}\label{eq:8}
f_0^i(x,a), f_1^i(x,a), ... ,f_p^i(x,a),\,\,  i= 1, ... ,n.
\end{eqnarray}
Let us define the one-parameter family G of approximate transformations
\begin{eqnarray}\label{eq:9}
\bar{x}^i \approx f_0^i(x,a)+\varepsilon f_1^i(x,a)+\cdots+\varepsilon^p f_p^i(x,a),\,\, i= 1,\cdots ,n
\end{eqnarray}
of points $x = (x^1,\cdots,x^n )\in \mathbb{R}^n$  into points $\bar{x}= (\bar{x}^1,\cdots,\bar{x}^n)\in\mathbb{R}^n$ as the class of
invertible transformations
\begin{eqnarray}\label{eq:10}
\bar{x}= f(x,a,\epsilon),
\end{eqnarray}
with vector-functions $f=(f^1,\cdots,f^n)$ such that 
\begin{eqnarray}\label{eq:11}
f^i(x,a,\epsilon)\approx f_0^i(x,a) + \epsilon f_1^i(x, a) +\cdots+\varepsilon^p f_p^i~(x,a),\,\,  i = 1,\cdots ,n.
\end{eqnarray}
Here $a$ is a real parameter, and the following condition is imposed:
\begin{eqnarray}\label{eq:12}
f(x,0,\epsilon) \approx x.
\end{eqnarray}
\paragraph{Definition}
The set of transformations (\ref{eq:9}) is called a one-parameter approximate
transformation group if
\begin{eqnarray}\label{eq:13}
f(f(x,a,\varepsilon),b,\epsilon) \approx f(x,a+b,\varepsilon)
\end{eqnarray}
for all transformations (\ref{eq:10}).
\paragraph{Definition}
Let $G$ be a one-parameter approximate transformation group:
\begin{eqnarray}\label{eq:14}
\bar{z}^i \approx f(z,a,\varepsilon) \equiv f_0^i(z,a)+\varepsilon f_1^i (z,a),\,\,   i =1,\cdots,N.
\end{eqnarray}
An approximate equation
\begin{eqnarray}\label{eq:15}
F(z,\varepsilon)\equiv F_0(z)+\varepsilon F_1(z)\approx 0
\end{eqnarray}
is said to be approximately invariant with respect to $G$, or admits $G$ if
\begin{eqnarray}\label{eq:16}
F(\bar{z},\varepsilon) \approx F(f(z,a,\varepsilon),\varepsilon)=o(\varepsilon)
\end{eqnarray}
whenever $z = (z^l,\cdots,z^N)$ satisfies Eq. (\ref{eq:15}).
If $z = (x,u,u_{(1)},\cdots,u_{(k)})$ then (\ref{eq:15}) becomes an approximate differential equation
of order $k$, and $G$ is an approximate symmetry group of the differential equation.
\paragraph{Theorem}
Eq. (\ref{eq:15}) is approximately invariant under the approximate transformation
group (\ref{eq:14}) with the generator
\begin{eqnarray}\label{eq:17}
X=X_0+\varepsilon X_1\equiv \xi_0^i(z)\frac{\partial}{\partial z^i}+\varepsilon \xi_1^i\frac{\partial}{\partial z^i},
\end{eqnarray}
if and only if 
\begin{eqnarray}\label{eq:18}
[X^{(k)}F(z,\varepsilon)]_{F \approx0}=o(\varepsilon),
\end{eqnarray}
or
\begin{eqnarray}\label{eq:19}
[X_0^{(k)}F_0(z)+\varepsilon(X_1^{(k)}F_0(z)+X_0^{(k)}F_1(z))]_{(2.5)}=o(\varepsilon).
\end{eqnarray}
where $X^{(k)}$ is the prolongation of $X$  of order $k$. 
The operator (\ref{eq:17}) satisfying Eq. (\ref{eq:19}) is called an infinitesimal approximate
symmetry of, or an approximate operator admitted by Eq. (\ref{eq:15}). Accordingly, Eq.
(\ref{eq:19}) is termed the determining equation for approximate symmetries. 
\paragraph{Theorem}
If Eq. (\ref{eq:15}) admits an approximate tramformation group with the
generator $X=X_0+ \varepsilon X_1$,  where $X_0\neq 0$, then the operator
\begin{eqnarray}\label{eq:20}
X_0=\xi_0^i(z)\frac{\partial}{\partial z^i}
\end{eqnarray}
is an exact symmetry ofthe equation
\begin{eqnarray}\label{eq:21}
F_0(z) =0.
\end{eqnarray}
\paragraph{Definition}
Eqs. (\ref{eq:21}) and (\ref{eq:15}) are termed an unperturbed equation and a
perturbed equation, respectively. Under the conditions of Theorem 2.3, the operator
$X_0$ is called a stable symmetry of the unperturbed equation (\ref{eq:21}). The corresponding
approximate symmetry generator $X=X_0+ \varepsilon X_1$ for the perturbed equation (\ref{eq:15}) 
 is called a deformation of the infinitesimal symmetry $X_0$ of Eq. (\ref{eq:21}) caused
by the perturbation $\varepsilon F_1(z)$. In particular, if the most general symmetry Lie algebra
of Eq. (\ref{eq:21}) is stable, we say that the perturbed equation (\ref{eq:15}) inherits the
symmetries of the unperturbed equation.
\section{Approximate symmetries of the perturbed Harry Dym equation}
Consider the perturbed Harry Dym equation 
\begin{eqnarray*}
u_t+\frac{1}{2}u^3u_{xxx}+\varepsilon u_x=0.
\end{eqnarray*}
By applying the method of approximate transformation groups, we provide the infinitesimal approximate symmetries (\ref{eq:17}) for the perturbed Harry Dym equation (\ref{eq:22}).
\subsection{Exact symmetries}
Let us consider the approximate group generators in the form
\begin{eqnarray}\label{eq:23}
X =X_0 +\varepsilon X_1 = (\xi_0+\varepsilon\xi_1)\frac{\partial}{\partial x} +(\tau_0+\varepsilon\tau_1)\frac{\partial}{\partial t}+
(\phi_0+\varepsilon\phi_1)\frac{\partial}{\partial u} 
\end{eqnarray}
where $r\xi_i, \tau_i$ and $\phi_i$  for $i=0,1$  are unknown functions of$x, t$  and $u$.
Solving the determining equation
\begin{eqnarray}\label{eq:24}
X_0^{(3)}(u_t-\frac{1}{2}u^3u_{xxx})\mid_{u_t-\frac{1}{2}u^3u_{xxx}=0}=0,
\end{eqnarray}
 for the exact symmetries $X_0$ of
the unperturbed equation, we obtain
\begin{eqnarray}\label{eq:25}
\xi_0&=&(A_1+A_2x+\frac{A_3}{2}x^2),\nonumber \\ 
\tau_0&=&(A_4+3A_5t),\\
\phi_0&=&(A_2-\frac{1}{3A_5}+xA_3)u,\nonumber
\end{eqnarray}
where $A_1 ,\cdots, A_5$ are arbitrary constants. Hence, 
\begin{eqnarray}\label{eq:26}
X_0=(A_1+A_2x+\frac{A_3}{2}x^2)\frac{\partial}{\partial x}+(A_4+3A_5t)\frac{\partial}{\partial t}+(A_2-\frac{1}{3A_5}+xA_3)u)\frac{\partial}{\partial u}
\end{eqnarray}
Therefore, the unperturbed Harry Dym equation, admits the five-dimensional Lie algebra with the basis
\begin{eqnarray}\label{eq:27}
\begin{array}{l}
\displaystyle X_0^1=\frac{\partial}{\partial x}, \\[2mm]
\displaystyle X_0^2=\frac{\partial}{\partial t}, \\[2mm]
\displaystyle X_0^3=x\frac{\partial}{\partial x}+u\frac{\partial}{\partial u},
\end{array} \qquad 
\begin{array}{l}
\displaystyle X_0^4=3t\frac{\partial}{\partial t}-u\frac{\partial}{\partial u}, \\[2mm]
\displaystyle X_0^5=2x^2\frac{\partial}{\partial x}+xu\frac{\partial}{\partial u}.
\end{array}
\end{eqnarray}
\subsection{Approximate symmetries}
At first, we need to determine the auxiliary finction $H$ by vitue of Eqs. (\ref{eq:18}), (\ref{eq:19}) and (\ref{eq:15}), i.e., by the equation
\begin{eqnarray}\label{eq:28}
H=\frac{1}{\varepsilon}[X_0^{(k)}(F_0(z)+\varepsilon F_1(z))\mid_{F_0(z)+\varepsilon F_1(z)=0}]
\end{eqnarray}
Substituting the expression (\ref{eq:26}) of the generator $X_0$ into Eq. (\ref{eq:28})
we obtain the auxiliary function
\begin{eqnarray}\label{eq:29}
H=u_x(A_5-A_2)+A_3(u-xu_x)
\end{eqnarray}
Now, calculate the operators $X_1$ by solving the inhomogeneous determining equation
for deformations:
\begin{eqnarray}\label{eq:30}
X_1^{(k)} F_0(z)\mid_{F_0(z)=0}+H=0.
\end{eqnarray}
So, the above determinig equation for this equation is written as 
\begin{eqnarray}\label{eq:31}
X_1^{(3)}(u_t+\frac{1}{2}u^3u_{xxx})\mid_{u_t+\frac{1}{2}u^3u_{xxx}=0}+u_x(A_5-A_2)+A_3(u-xu_x)=0, 
\end{eqnarray}
solving the determining equation yields,
\begin{eqnarray}\label{eq:32}
\xi_1&=&(A_5-A_2)t-A_3xt+C_4x-C_5+\frac{C_3}{2}x^2, \nonumber \\
\tau_1&=&(C_1t+C_2), \\
\phi_1&=&(-A_3t+C_4+C_3x+\frac{C_1}{3})u, \nonumber 
\end{eqnarray} 
where $C_1 ,\cdots, C_5$ are arbitrary constants. 

Thus, we derive the following approximate symmetries of the perturvbed Harry Dym equation:
\begin{eqnarray}\label{eq:33}
\begin{array}{l}
\displaystyle \mathbf{v_1}=\frac{\partial}{\partial x}, \\[2mm]
\displaystyle \mathbf{v}_2=\frac{\partial}{\partial t}, \\[2mm]
\displaystyle \mathbf{v}_3=x\frac{\partial}{\partial x}+u\frac{\partial}{\partial u},\\[2mm]
\displaystyle \mathbf{v}_4=3t\frac{\partial}{\partial t}-u\frac{\partial}{\partial u}, 
\end{array} \qquad  
\begin{array}{l}
\displaystyle \mathbf{v}_5=2x^2\frac{\partial}{\partial x}+xu\frac{\partial}{\partial u},\\[2mm]
\displaystyle \mathbf{v}_6=\varepsilon\frac{\partial}{\partial x}, \\[2mm]
\displaystyle \mathbf{v}_7=\varepsilon\frac{\partial}{\partial t}, \\[2mm]
\displaystyle \mathbf{v}_8=\varepsilon (x\frac{\partial}{\partial x}+u\frac{\partial}{\partial u}),
\end{array} \qquad 
\begin{array}{l}
\displaystyle \mathbf{v}_9=\varepsilon(3t\frac{\partial}{\partial t}-u\frac{\partial}{\partial u}), \\[2mm]
\displaystyle \mathbf{v}_{10}=\varepsilon(2x^2\frac{\partial}{\partial x}+xu\frac{\partial}{\partial u}).
\end{array}
\end{eqnarray}
The following table of commutators, evaluated in the first-order of precision,
shows that the operators (\ref{eq:33}) span an ten-dimensional approximate Lie algebra
, and hence generate an ten-parameter approximate transformations group. 
\begin{table}[h]
\caption{Approximate Commutators of approximate symmetry of perturbed Harry Dym equation}
\begin{eqnarray*}
\begin{array}{c|cccccccccc} &\mathbf{v_1}&\mathbf{v}_2&\mathbf{v}_3&\mathbf{v}_4&\mathbf{v}_5&\mathbf{v}_6&\mathbf{v}_7&\mathbf{v}_8&\mathbf{v}_9&\mathbf{v}_{10}\\\hline\ &&&&&& \\[-3mm]
\mathbf{v_1}&0   &0   &\mathbf{v_1}   &0 &2\mathbf{v}_3 &0 &0 &\mathbf{v}_6 &0  &2\mathbf{v}_8\\
\mathbf{v}_2&0   &0   &0   &12\mathbf{v}_2    &0   &0   &0   &0   &3\mathbf{v}_7   &0\\
\mathbf{v}_3&-\mathbf{v}_1   &0   &0   &0   &\mathbf{v}_5   &-\mathbf{v}_6   &0   &0   &0   &\mathbf{v}_{10}   \\
\mathbf{v}_4&0   &-12\mathbf{v}_2   &0 &0   &0   &0   &-3\mathbf{v}_7   &0   &0   &0\\
\mathbf{v}_5&-2\mathbf{v}_3 &0   &-\mathbf{v}_5   &0    &0   &-2\mathbf{v}_8   &0   &-\mathbf{v}_{10}   &0   &0\\
\mathbf{v}_6&0   &0 &\mathbf{v}_6   &0   &2\mathbf{v}_8 &0   &0   &0   &0   &0\\
\mathbf{v}_7&0   &0   &0   &3\mathbf{v}_7  &0   &0  &0   &0   &0   &0  \\
\mathbf{v}_8&-\mathbf{v}_6   &0   &0   &0   &\mathbf{v}_{10}  &0   &0   &0   &0   &0\\
\mathbf{v}_9&0   &-3\mathbf{v}_7   &0   &0   &0  &0   &0   &0   &0  &0\\
\mathbf{v}_{10}&-2\mathbf{v}_8   &0   &-\mathbf{v}_{10}  &0   &0  &0   &0   &0   &0   &0
\end{array}
\end{eqnarray*}
\end{table}
\paragraph{Remark.} Equations (\ref{eq:33}) show that all symmetries (\ref{eq:27}) of Eq. (\ref{eq:1}) are
stable. Hence, the perturbed equation (\ref{eq:22}) inherits the symmetries of the unperturbed
equation (\ref{eq:1}).
\section{The structure of the Lie algebra of symmetries}
In this section, we determine the structure of the Lie algebra of symmetries of the perturbed 
Harry Dym equation. 
The Lie algebra $\mathfrak{g}$ is non-solvable, since 
\begin{eqnarray}\label{eq:34}
\mathfrak{g}^{(1)}&=&[\mathfrak{g},\mathfrak{g}]=\mathrm{Span}_\mathbb{R}\{\mathbf{v}_1, \mathbf{v}_2, \mathbf{v}_3, \mathbf{v}_5,
 \mathbf{v}_6,\mathbf{v}_7, \mathbf{v}_8,\mathbf{v}_{10}\}\nonumber\\
\mathfrak{g}^{(2)}&=&[\mathfrak{g}^{(1)},\mathfrak{g}^{(1)}]=\mathrm{Span}_\mathbb{R}\{\mathbf{v}_1, \mathbf{v}_3, \mathbf{v}_5,
 \mathbf{v}_6, \mathbf{v}_8,\mathbf{v}_{10}\}\\
 \mathfrak{g}^{(3)}&=&[\mathfrak{g}^{(2)},\mathfrak{g}^{(2)}]=\mathfrak{g}^{(2)}\nonumber
\end{eqnarray}
The Lie algebra $\mathfrak{g}$ admits a Levi decomposition as the following semi-direct product 
$\mathfrak{g}=r\propto s$, where 
\begin{eqnarray}\label{eq:35}
r=\mathrm{Span}_\mathbb{R}\{\mathbf{v}_2, \mathbf{v}_4, \mathbf{v}_6, \mathbf{v}_7, \mathbf{v}_8, \mathbf{v}_9, \mathbf{v}_{10}\}
\end{eqnarray}
is the radical of $\mathfrak{g}$ (the largest solvable ideal contained in $\mathfrak{g}$) and 
\begin{eqnarray}\label{eq:36}
s=\mathrm{Span}_\mathbb{R}\{\mathbf{v}_1, \mathbf{v}_3, \mathbf{v}_5\}
\end{eqnarray}
is a semi-simple subalgebra of $\mathfrak{g}$.

The radical $r$ is solvable with the following chain of ideals
\begin{eqnarray}\label{eq:37}
r^{(1)}\supset r^{(2)}\supset r^{(3)}=\{0\},
\end{eqnarray}
where
\begin{eqnarray}\label{eq:38}
r^{(1)}&=&\mathrm{Span}_\mathbb{R}\{\mathbf{v}_2, \mathbf{v}_4, \mathbf{v}_6, \mathbf{v}_7, \mathbf{v}_8, \mathbf{v}_9, \mathbf{v}_{10}\}, \nonumber \\
r^{(2)}&=&\mathrm{Span}_\mathbb{R}\{\mathbf{v}_2,\mathbf{v}_7\}.
\end{eqnarray}
The semi-simple subalgebra $s$ of $\mathfrak{g}$ is isomorphic to the Lie algebra $A_{3,8}$ of the classification of three dimensional Lie algebras in \cite{14}, by the following isomorphism.
\begin{eqnarray}\label{eq:39}
\mathfrak{T}:\{\mathbf{v}_1,\mathbf{v}_3, \mathbf{v}_5 \}\rightarrow \{\mathbf{v}_1, -\mathbf{v}_2, -\mathbf{v}_3\}
\end{eqnarray}
\section{Optimal system for perturbed Harry Dym equation}
\paragraph{Definition} Let $G$ be a Lie group. An optimal system of $s$-parameter
subgroups is a list of conjugacy inequivalent $s$-parameter subgroups with the
property that any other subgroup is conjugate to precisely one subgroup in
the list. Similarly, a list of $s$-parameter subalgebras forms an optimal system if
every $s$-parameter subalgebra of $\mathfrak{g}$ is equivalent to a unique member of the list
under some element of the adjoint representation: $ \tilde{\mathfrak{h}}=\mathrm{ Ad} g(\mathfrak{h})),\, g\in G$.
\paragraph{Proposition} Let $H$ and  $\tilde{H}$ be connected, $s$-dimensional Lie subgroups of the
Lie group $G$ with corresponding Lie subalgebras $\mathfrak{h}$ and $\tilde{\mathfrak{h}}$ of the Lie algebra $\mathfrak{g}$
of $G$. Then $\tilde{H} = gHg^{-1}$ are conjugate subgroups if and only if $\tilde{\mathfrak{h}} =\mathrm{Ad}g(\mathfrak{h}) $are
conjugate subalgebras.(Proposition 3.7 of \cite{12})

Actually, the proposition  says that the problem of finding an optimal system of
subgroups is equivalent to that of finding an optimal system of subalgebras. 
For one-dimensional subalgebras, this classification problem is essentially
the same as the problem of classifying the orbits of the adjoint representation,
since each one-dimensional subalgebra is determined by a nonzero vector in
$\mathfrak{g}$. To compute the adjoint representation we use the Lie series 
\begin{eqnarray}\label{eq:40}
\mathrm{Ad}(\mathrm{exp}(\mu \mathbf{v}_i))\mathbf{v}_j=\mathbf{v}_j-\mu[\mathbf{v}_i,\mathbf{v}_j]+\frac{\mu^2}{2}
[[\mathbf{v}_i,[\mathbf{v}_i,\mathbf{v}_j]]-\cdots.
\end{eqnarray}
where $[\mathbf{v}_i,\mathbf{v}_j], \, i,j=1,\cdots, 10$ is the commutator for the Lie algebra and $\mu$ is a parameter.
In this manner, we construct the table with the $(i, j)$-th entry indicating $\mathrm{Ad}(\mathrm{exp}(\mu \mathbf{v}_i))\mathbf{v}_j$.
\begin{table}[h]
\caption{Adjoint representation of approximate symmetry of the perturbed Harry Dym equation}
\begin{eqnarray*}
\begin{array}{c|cccccccccc}\mathrm{Ad}
&\mathbf{v}_1&\mathbf{v}_2&\mathbf{v}_3&\mathbf{v}_4&\mathbf{v}_5 \\\hline\ &&&&&& \\[-3mm]
\mathbf{v}_1 & \mathbf{v}_1 & \mathbf{v}_2  &\mathbf{v}_3-\mu \mathbf{v}_1   &\mathbf{v}_4 &\mathbf{v}_5-2\mu\mathbf{v}_3+\mu^2\mathbf{v}_1 \\
\mathbf{v}_2 & \mathbf{v}_1   &\mathbf{v}_2  &\mathbf{v}_3   &\mathbf{v}_4-12\mu \mathbf{v}_2   &\mathbf{v}_5 \\
\mathbf{v}_3 & e^\mu\mathbf{v}_1   & \mathbf{v}_2  &\mathbf{v}_3   &\mathbf{v}_4   &e^{-\mu}\mathbf{v}_5   \\
\mathbf{v}_4 & \mathbf{v}_1 &e^{12\mu}\mathbf{v}_2   &\mathbf{v}_3 &\mathbf{v}_4   &\mathbf{v}_5  \\
\mathbf{v}_5&\mathbf{v}_1+2\mu  \mathbf{v}_3+\mu^2 \mathbf{v}_5&\mathbf{v}_2 &\mathbf{v}_3+\mu \mathbf{v}_5   &\mathbf{v}_4    &\mathbf{v}_5  \\
\mathbf{v}_6& \mathbf{v}_1  &\mathbf{v}_2 &\mathbf{v}_3-\mu \mathbf{v}_6   &\mathbf{v}_4   &\mathbf{v}_5-2\mu\mathbf{v}_8 \\
\mathbf{v}_7&\mathbf{v}_1   &\mathbf{v}_2   &\mathbf{v}_3   &\mathbf{v}_4-3\mu\mathbf{v}_7  &\mathbf{v}_5   \\
\mathbf{v}_8&\mathbf{v}_1+\mu\mathbf{v}_6   &\mathbf{v}_2   &\mathbf{v}_3   &\mathbf{v}_4 &\mathbf{v}_5-\mu\mathbf{v}_{10} \\
\mathbf{v}_9&\mathbf{v}_1  &\mathbf{v}_2+3\mu\mathbf{v}_7   &\mathbf{v}_3   &\mathbf{v}_4   &\mathbf{v}_5  \\
\mathbf{v}_{10}&\mathbf{v}_1+2\mu\mathbf{v}_8& \mathbf{v}_2& \mathbf{v}_3+\mu\mathbf{v}_{10}& \mathbf{v}_4& \mathbf{v}_5
\end{array}
\end{eqnarray*}
\begin{eqnarray*}
\begin{array}{c|cccccccccc}\mathrm{Ad}
&\mathbf{v}_6  & \mathbf{v}_7&\mathbf{v}_8&\mathbf{v}_9&\mathbf{v}_{10}\\\hline\ &&&&&& \\[-3mm]
\mathbf{v}_1  & \mathbf{v}_6 & \mathbf{v}_7 &\mathbf{v}_8-\mu \mathbf{v}_6 &\mathbf{v}_9  &\mathbf{v}_{10}-2\mu \mathbf{v}_8+\mu^2 \mathbf{v}_6\\
\mathbf{v}_2  & \mathbf{v}_6   &\mathbf{v}_7   &\mathbf{v}_8   &\mathbf{v}_9-3\mu \mathbf{v}_7   &\mathbf{v}_{10}\\
\mathbf{v}_3 & e^\mu\mathbf{v}_6   &\mathbf{v}_7   &\mathbf{v}_8  &\mathbf{v}_9   &e^{-\mu}\mathbf{v}_{10}   \\
\mathbf{v}_4 & \mathbf{v}_6   &e^{3\mu}\mathbf{v}_7   &\mathbf{v}_8   &\mathbf{v}_9  &\mathbf{v}_{10}\\
\mathbf{v}_5 & \mathbf{v}_6+2\mu \mathbf{v}_8 +\mu^2 \mathbf{v}_{10}   &\mathbf{v}_7  &\mathbf{v}_8+\mu+\mathbf{v}_{10}   &\mathbf{v}_9  &\mathbf{v}_{10}\\
\mathbf{v}_6 & \mathbf{v}_6   &\mathbf{v}_7   &\mathbf{v}_8   &\mathbf{v}_9   &\mathbf{v}_{10}\\
\mathbf{v}_7 & \mathbf{v}_6  &\mathbf{v}_7   &\mathbf{v}_8   &\mathbf{v}_9   &\mathbf{v}_{10}  \\
\mathbf{v}_8 & \mathbf{v}_6   &\mathbf{v}_7  &\mathbf{v}_8   &\mathbf{v}_9   &\mathbf{v}_{10}\\
\mathbf{v}_9 & \mathbf{v}_6   &\mathbf{v}_7   &\mathbf{v}_8   &\mathbf{v}_9  &\mathbf{v}_{10}\\
\mathbf{v}_{10} & \mathbf{v}_6& \mathbf{v}_7& \mathbf{v}_8& \mathbf{v}_9& \mathbf{v}_{10}
\end{array}
\end{eqnarray*}
\end{table}

\paragraph{Theorem}
An optimal system of one-dimensional approximate Lie algebras of the perturbed Harry Dym equation is provided by 
\begin{table}[h]
\caption{One-dimensional optimal system of the perturbed Harry Dym equation }
\begin{eqnarray*}
&& \hspace{-17mm}
\begin{array}{l}
\mathbf{v}^1=\mathbf{v}_8, \\
\mathbf{v}^2=\mathbf{v}_7+a\mathbf{v}_8, \\
\mathbf{v}^3=\mathbf{v}_6+\mathbf{v}_8,
\end{array} \qquad 
\begin{array}{l}
\mathbf{v}^4=\mathbf{v}_6-\mathbf{v}_7+\mathbf{v}_8,  \\
\mathbf{v}^5=\mathbf{v}_6+\mathbf{v}_7+\mathbf{v}_8,\\
\mathbf{v}^6=\mathbf{v}_2+a\mathbf{v}_8, 
\end{array} \qquad 
\begin{array}{l}
\mathbf{v}^7=\mathbf{v}_2-\mathbf{v}_6+a\mathbf{v}_8, \\
\mathbf{v}^8=\mathbf{v}_2+\mathbf{v}_6+a\mathbf{v}_8, \\
\mathbf{v}^9=\mathbf{v}_1+a\mathbf{v}_2+b\mathbf{v}_7
\end{array} \\
\mathbf{v}^{10}\!\!&=&\!\!a\mathbf{v}_1+b\mathbf{v}_2+\mathbf{v}_5+c\mathbf{v}_6+d\mathbf{v}_7,\\
\mathbf{v}^{11}\!\!&=&\!\!a\mathbf{v}_1+b\mathbf{v}_2+\mathbf{v}_3+c\mathbf{v}_5+d\mathbf{v}_7+e\mathbf{v}_8,\nonumber\\
\mathbf{v}^{12}\!\!&=&\!\!a\mathbf{v}_1+b\mathbf{v}_3+\mathbf{v}_4+c\mathbf{v}_5+d\mathbf{v}_6+e\mathbf{v}_8,\nonumber\\
\mathbf{v}^{13}\!\!&=&\!\!a\mathbf{v}_1+b\mathbf{v}_3+c\mathbf{v}_4+d\mathbf{v}_5+e\mathbf{v}_6+f\mathbf{v}_8+\mathbf{v}_9,\nonumber\\
\mathbf{v}^{14}\!\!&=&\!\!a\mathbf{v}_1-\mathbf{v}_2+b\mathbf{v}_3+c\mathbf{v}_4+d\mathbf{v}_5+e\mathbf{v}_6+f\mathbf{v}_8+\mathbf{v}_9,\nonumber\\
\mathbf{v}^{15}\!\!&=&\!\!a\mathbf{v}_1+\mathbf{v}_2+b\mathbf{v}_3+c\mathbf{v}_4+d\mathbf{v}_5+e\mathbf{v}_6+f\mathbf{v}_8+\mathbf{v}_9,\nonumber\\
 \mathbf{v}^{16}\!\!&=&\!\!a\mathbf{v}_1+\mathbf{v}_2+b\mathbf{v}_3+c\mathbf{v}_4+d\mathbf{v}_5+e\mathbf{v}_6+f\mathbf{v}_8+\mathbf{v}_9.\nonumber
\end{eqnarray*}
\end{table}

Proof. Consider the approximate symmetry algebra $\mathfrak{g}$ of the unperturbed Harry Dym
equation, whose adjoint representation was determined in the table 2.
Given a nonzero vector 
\begin{eqnarray*}
\mathbf{v}=\sum_{i=1}^{10} a_i\mathbf{v_i},
\end{eqnarray*}
our task is to simplify as many of the coefficients ai as possible through
judicious applications of adjoint maps to $\mathbf{v}$.

Suppose first that $a_{10}\neq 0$. Scaling $\mathbf{v}$ if necessary, we can assume that
$a_{10} = 1$. Referring to table (3.24), if we act on such a $\mathbf{v}$ by
\begin{eqnarray*}
\mathbf{v}'&=&\mathrm{Ad}(\mathrm{exp}(\frac{a_8}{2} \mathbf{v}_8))\mathbf{v}\\&=&
a_1^{'} \mathbf{v}_1+a_2\mathbf{v}_2+a_3^{'} \mathbf{v}_3+a_4\mathbf{v}_4+a_5\mathbf{v}_5+a_6^{'} \mathbf{v}_6+a_7\mathbf{v}_7+a_9\mathbf{v}_9+\mathbf{v}_{10},
\end{eqnarray*}
we can make the coefficient of $a_8$ vanish. 
The remaining one-dimensional subalgebras are spanned by vectors of the above form with $a_{10}=0$. If $a_9\neq 0$, we scale to make 
$a_9=1$, and then act on $\mathbf{v}$ to cancel the coefficient of $a_7$ as follows:
\begin{eqnarray*}
\mathbf{v}'&=&\mathrm{Ad}(\mathrm{exp}(\frac{a_7}{3}  \mathbf{v}_2))\mathbf{v}\\&=&
a_1^{'} \mathbf{v}_1+a_2\mathbf{v}_2+a_3^{'} \mathbf{v}_3+a_4\mathbf{v}_4+a_5\mathbf{v}_5+a_6\mathbf{v}_6+a_8\mathbf{v}_8+\mathbf{v}_9.
\end{eqnarray*}
We can further act on $\mathbf{v}'$ by the group generated by $\mathbf{v}_4$; this has the net effect of
scaling the coefficients of $\mathbf{v}_2$:
\begin{eqnarray*}
\mathbf{v}''&=&\mathrm{Ad}(\mathrm{exp}(\mu \mathbf{v}_4))\mathbf{v}\\&=& a_1 \mathbf{v}_1+ e^{12\mu}a_2 \mathbf{v}_2+
a_3 \mathbf{v}_3+a_4 \mathbf{v}_4+a_5 \mathbf{v}_5+a_6\mathbf{v}_6+a_8 \mathbf{v}_8+\mathbf{v}_9.
\end{eqnarray*}
So, depending on the sign of $a_2$, we can make the coefficient of  $\mathbf{v}_2$  either $+1, -1$ or $0$.
If $a_{10}=a_9=0$ and $a_4\neq 0$, we scale to make $a_4=1$. So, the non-zero vector $ \mathbf{v}$ is equivalent to $ \mathbf{v}'$ 
under adjoint maps:
\begin{eqnarray*}
 \mathbf{v}'&=&\mathrm{Ad}(\mathrm{exp}(\frac{a_7}{3} \mathbf{v}_7))\circ\mathrm{Ad}(\mathrm{exp}(\frac{a_2}{12}  \mathbf{v}_2))\mathbf{v}\\&=&
a_1\mathbf{v}_1+a_3\mathbf{v}_3+\mathbf{v}_4+a_5\mathbf{v}_5+a_6\mathbf{v}_6+a_8\mathbf{v}_8
\end{eqnarray*}  
If $a_{10}=a_9=a_4=0$ and $a_3\neq 0$, by scaling $\mathbf{v}$, we can assume that $a_3=1$. Referring to the table, if we act on 
such a $\mathbf{v}$ by the following adjoint map, we can arrange that the coefficients of $a_6$ vanish. 
\begin{eqnarray*}
\mathbf{v}'&=&\mathrm{Ad}(\mathrm{exp}(a_6\mathbf{v}_6))\mathbf{v}\\&=&
a_1 \mathbf{v}_1+a_2 \mathbf{v}_2+\mathbf{v}_3+a_5\mathbf{v}_5+a_6 \mathbf{v}_6+a_7\mathbf{v}_7+a_8^{'}\mathbf{v}_8.
\end{eqnarray*}
If $a_{10}=a_9=a_4=a_3=0$ and $a_5\neq 0$, we scale to make $a_5=1$.
Thus, $\mathbf{v}$ is equivalent to $\mathbf{v}'$ under the adjoint representations:
\begin{eqnarray*}
\mathbf{v}'&=&\mathrm{Ad}(\mathrm{exp}(\frac{a_8}{2} \mathbf{v}_6))\mathbf{v}\\&=&
a_1\mathbf{v}_1+a_2\mathbf{v}_2+\mathbf{v}_5+a_6\mathbf{v}_6+a_7\mathbf{v}_7.
\end{eqnarray*} 
If $a_{10}=a_9=a_4=a_3=a_5=0$ and $a_1\neq 0$, we scale to make $a_1=1$. So, we can make the coefficients of $a_6, a_8$ 
zero by using the following adjoint maps:
\begin{eqnarray*}
\mathbf{v}'&=&\mathrm{Ad}(\mathrm{exp}(-\frac{a_8}{2}\mathbf{v}_{10}))\circ\mathrm{Ad}(\mathrm{exp}(-a_6\mathbf{v}_8))\mathbf{v}\\
&=&\mathbf{v}_1+a_2\mathbf{v}_2+a_7\mathbf{v}_7.\nonumber
\end{eqnarray*}
If $a_{10}=a_9=a_4=a_3=a_5=a_1=0$ and $a_2\neq 0$, by scaling $\mathbf{v}$, we can assume that $a_2=1$. Therefore,  we 
can arrange that the coefficients of $a_7$ vanish by simplifying 
the non-zero vector $\mathbf{v}$ as follows:
\begin{eqnarray*}
\mathbf{v}'&=&\mathrm{Ad}(\mathrm{exp}(-\frac{a_7}{3}\mathbf{v}_9))\mathbf{v}\\
&=&\mathbf{v}_2+a_6\mathbf{v}_6+a_8\mathbf{v}_8,\nonumber
\end{eqnarray*}
We can further act on $\mathbf{v}'$ by the group generated by $\mathbf{v}_3$;
\begin{eqnarray*}
\mathbf{v}''&=&\mathrm{Ad}(\mathrm{exp}(\mu \mathbf{v}_3)\mathbf{v}\\
&=&\mathbf{v}_2+ e^{\mu}a_6 \mathbf{v}_6+
a_8\mathbf{v}_8.
\end{eqnarray*}
So, depending on the sign of $a_6$, we can make the coefficient of  $\mathbf{v}_6$  either $+1, -1$ or $0$.
If $a_{10}=a_9=a_4=a_3=a_5=a_1=a_2=0$ and $a_6\neq 0$, by scaling $\mathbf{v}$, we can assume that $a_6=1$.  
We can act on such a $\mathbf{v}$ by the group generated by $\mathbf{v}_4$;
So, depending on the sign of $a_7$, we can make the coefficient of  $\mathbf{v}_7$  either $+1, -1$ or $0$.
The remaining cases $a_{10}=a_9=a_4=a_3=a_5=a_2=a_1=a_3=0$ and $a_7\neq 0$,
no further simplifications are possible.
The last remaining cace occurs when $a_{10}=a_9=a_4=a_3=a_5=a_1=a_2=a_4=a_6=a_7=0$ and $a_8\neq 0$, for which our
earlier simplifications were unnecessary. Since, the only remaining vectors are
the multiples of $\mathbf{v}_8$, on which the adjoint representation acts trivially. 
\section{Approximately differential  invariants for the perturbed Harry Dym equation}
In this section, we compute some approximately differential invariants of the  perturbed Harry Dym equation 
with respect to the optimal system.
Consider the operator $\mathbf{v}^2$. 
To determine the independent invariants $I$, we need to solve the first order partial differential equation
\begin{eqnarray}\label{eq:41}
(\varepsilon \frac{\partial}{\partial t}+a\varepsilon x \frac{\partial}{\partial x}+a\varepsilon u \frac{\partial}{\partial u})(I(x,t,u))=0,
\end{eqnarray}
that is
\begin{eqnarray}\label{eq:42}
\varepsilon \frac{\partial I}{\partial t}+a\varepsilon x \frac{\partial I}{\partial x}+a\varepsilon u \frac{\partial I}{\partial u}=0,
\end{eqnarray}
which is a first order homogeneous PDE. The solution can be found by integrating the corresponding characteristic system 
of ordinary differential equation, which is
\begin{eqnarray}\label{eq:43}
 \frac{d x}{a\varepsilon x}= \frac{d t}{\varepsilon}=\frac{d u}{a\varepsilon u}
\end{eqnarray}
Hence, the independent approximately differntial invariants are as follows: 
\begin{eqnarray}\label{eq:44}
y=\frac{u}{x}, \quad v=\frac{\ln x-at}{a}.
\end{eqnarray}
In this manner, we investigate some independent approximately differential invariants with respect to the optimal system which 
are listed in Table 4. 

\begin{table}[h]
\caption{Approximately differential invariants for the perturbed Harry Dym
equation}
\begin{center}
\begin{tabular}{lcl} 
\mbox{Operator} && \mbox{Approximate Differential Invariants}\\[2mm] \hline
$\displaystyle \mathbf{v}_1 $ && $  \displaystyle t, u$\\[2mm]
$\displaystyle \mathbf{v}_2 $ && $ \displaystyle x,u$\\[2mm]
$\displaystyle \mathbf{v}_3 $ && $ \displaystyle t, \frac{u}{x}$\\[2mm]
$\displaystyle \mathbf{v}_4 $ && $  \displaystyle x, ut^{1/3}$\\[2mm]
$\displaystyle \mathbf{v}_5$ && $  \displaystyle t, \frac{u}{x^2}$\\[2mm]
$\displaystyle \mathbf{v}_7+a\mathbf{v}_8 $ && $  \displaystyle -\frac{\ln x}{a}+t, \frac{u}{x}$\\[2mm]
$\displaystyle \mathbf{v}_6+\mathbf{v}_8$ && $  \displaystyle t, \frac{u}{x+1}$\\[2mm]
$\displaystyle \mathbf{v}_6-\mathbf{v}_7+\mathbf{v}_8 $ && $ \displaystyle  \ln(x+1)+t, \frac{u}{x+1}$\\[2mm]
$\displaystyle \mathbf{v}_6+\mathbf{v}_7+\mathbf{v}_8 $ && $ \displaystyle  -\ln(x+1)+t, \frac{u}{x+1}$\\[2mm]
$\displaystyle \mathbf{v}_2+a\mathbf{v}_8 $ && $ \displaystyle  -\frac{\ln x}{a\varepsilon}+t,\frac{u}{x}$\\[2mm]
$\displaystyle \mathbf{v}_2-\mathbf{v}_6+a\mathbf{v}_8 $ && $ \displaystyle  -\frac{\ln (ax-1)}{a\varepsilon}+t, \frac{u}{ax-1}$\\[2mm]
$\displaystyle \mathbf{v}_2+\mathbf{v}_6+a\mathbf{v}_8 $ && $ \displaystyle  -\frac{\ln (ax+1)}{a\varepsilon}+t, \frac{u}{ax+1}$\\[2mm]
$\displaystyle \mathbf{v}_1+a\mathbf{v}_2+b\mathbf{v}_7 $ && $ \displaystyle  -b\varepsilon x-a\varepsilon+t, u$\\[2mm]
$\displaystyle a\mathbf{v}_1+ b\mathbf{v}_2+ \mathbf{v}_5+ c\mathbf{v}_6+ d\mathbf{v}_7$ && $\displaystyle  \frac{-d \varepsilon-b}{\sqrt{c\varepsilon +a}}\arctan\Big(\frac{x}{\sqrt{c\varepsilon +a}}\Big)+t, \frac{u}{x^2+c\varepsilon+a}$ \\[4mm] \hline
\end{tabular}
\end{center}
\end{table}
\section{Conclusions}
In this paper, we investigate the approximate symmetry of the perturbed Harry Dym equation and discuss on the structure of its Lie algebra.
Moreover, we compute  optimal system of one-dimensional approximate Lie algebras of the perturbed
 Harry Dym equation and derive some approximately differential invarints with respect 
to the generators of Lie algebra and optimal system.

\begin{footnotesize}
\begin{flushleft}
Mehdi Nadjafikhah\\
\textit{Iran University of Science and Technology, Narmak-16,
Tehran, Iran}\\
\textit{E-mail:} m\_nadjafikhah@iust.ac.ir
\end{flushleft}
\end{footnotesize}

\begin{footnotesize}
\begin{flushleft}
Parastoo Kabi-Nejad\\
\textit{Iran University of Science and Technology, Narmak-16,
Tehran, Iran}\\
\textit{E-mail:} parastoo\_kabinejad@iust.ac.ir
\end{flushleft}
\end{footnotesize}
\end{document}